\documentclass{jaa}

\usepackage{txfonts}
\usepackage{balance}
\usepackage{textcase}
\usepackage{graphicx}
\usepackage{cite}
\usepackage{amsmath,amssymb,amsfonts}
\usepackage{algorithmic}
\usepackage{float}
\usepackage{placeins}
\usepackage{listings}
\usepackage{verbatim}
\usepackage{textcomp}
\begin{document}\sloppy
\graphicspath{ {./images/} }
\title{An Improved Approach to Orbital Determination and Prediction of Near-Earth Asteroids: Computer Simulation, Modeling and Test Measurements}


\author{Muhammad Farae\textsuperscript{1*}, Cameron Woo\textsuperscript{1}, Anka Hu\textsuperscript{1}}

\affilOne{\textsuperscript{1}Department of Physics, University of Colorado Boulder, Boulder CO 80309 USA\\}


\twocolumn[{

\maketitle

\corres{muhammad.farae@gmail.com}


\begin{abstract}
In this article, theory-based analytical methodologies of astrophysics employed in the modern era are suitably operated alongside a test research-grade telescope to image and determine the orbit of a near-earth asteroid from original observations, measurements, and calculations. Subsequently, its intrinsic orbital path has been calculated including the chance it would likely impact Earth in the time ahead. More so specifically, this case-study incorporates the most effective, feasible, and novel Gauss's Method in order to maneuver the orbital plane components of a planetesimal, further elaborating and extending our probes on a selected near-earth asteroid (namely the 12538-1998 OH) through the observational data acquired over a six week period. Utilizing the CCD (Charge Coupled Device) snapshots captured, we simulate and calculate the orbit of our asteroid as outlined in quite detailed explanations. The uncertainties and deviations from the expected values are derived to reach a judgement whether our empirical findings are truly reliable and representative measurements by partaking a statistical analysis based systematic approach. Concluding the study by narrating what could have caused such discrepancy of findings in the first place, if any, measures are put forward that could be undertaken to improve the test-case for future investigations. Following the calculation of orbital elements and their uncertainties using Monte Carlo analysis, simulations were executed with various sample celestial bodies to derive a plausible prediction regarding the fate of Asteroid 1998 OH. Finally, the astrometric and photometric data, after their precise verification, were officially submitted to the Minor Planet Center: an organization hosted by the Center for Astrophysics, Harvard and Smithsonian and funded by NASA, for keeping track of the asteroid's potential trajectories.
\end{abstract}

 \vspace{20mm}}]
\pgrange{1--}
\setcounter{page}{1}
\lp{1}

\section{Introduction}

The Solar System, in its current state, is rife with planetesimals forming via the collision, spallation, and iterative fragmentation of larger planetary bodies. The paramount constituents of these minor celestial bodies are namely asteroids and comets, each with a uniquely defined orbit (Milone \& Wilson 2014). This study focuses specifically on NEAs (Near-Earth Asteroids) which have a perihelion less than 1.3 AU and an aphelion greater than or equal to 0.983 AU from the Sun (NASA 2007). An Earth-crossing asteroid is one which has an orbit intersecting the Earth's around the Sun, and a Mars-crosser is defined analogously.

As a result of the ubiquity of asteroids residing in our Solar System, tons of space debris enter the Earth's atmosphere each year, most of which are not massive enough to cause fatal destruction. However, some asteroids have potential to impact the earth with immense energy and cataclysmic out-turns (Chodas {\em et al.} 2001). With the exponential evolution of imaging technology and the surge in astronomical, and specifically orbital, research over the past years, enormous breakthroughs have been witnessed in the orbital determination of NEAs (Raol \& Sinha 1985; Farnocchia {\em et al.} 2013). In order to make the future behaviour of asteroids in the Solar System more predictable, multiple search campaigns have been established to map the night sky (el-Showk 2017; Yeomans {\em et al.} 1997). Our research work is a consequence of the aforementioned global mission astronomers worldwide are working towards solving: obtaining the orbital elements of asteroids and using those measurements to further expand our awareness of the night sky (Margot {\em et al.} 2002), thus enabling us to prepare for future asteroid impacts and mitigate them.  

\begin{figure}
\begin{center}
\includegraphics[width=18pc]{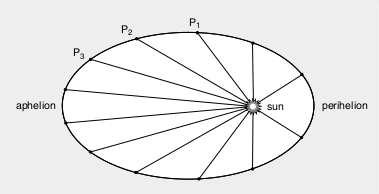}
\caption{Elliptical orbit of the asteroid and its three position vectors (Courtesy of: How Gauss Determined the Orbit of Ceres, pp. 35, 1997).}
\end{center}
\end{figure}

\begin{figure}
\begin{center}
\includegraphics[width=19pc]{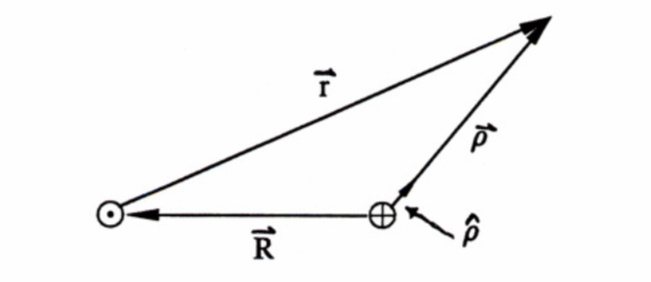}
\caption{
\label{fig2}
Summary of vectors (Earth-Sun, Earth-asteroid, and Sun-asteroid) involved in the orbit determination of a heliocentric asteroid.}
\end{center}
\end{figure}

Next the refined techniques for Orbital Determination (OD) are narrated. For the purpose of OD (Boulet 1991), Gauss's Method (https://en.wikipedia.org/wiki/Gauss\%27s\_method) was taken into account. Implementation and application of trajectory fitting necessitated two-fold crucial steps: (1) the preliminary orbit determination by calculating the position and velocity vectors of 1998 OH via Gauss's Method, (2) and finally, the computation of orbital elements (Richard 2008; NEODyS 2020; Asteroid (12538) 1998 OH, 2020) upon substitution of the results from step one.

\FloatBarrier
\begin{figure}
\begin{center}
\includegraphics[width=22pc]{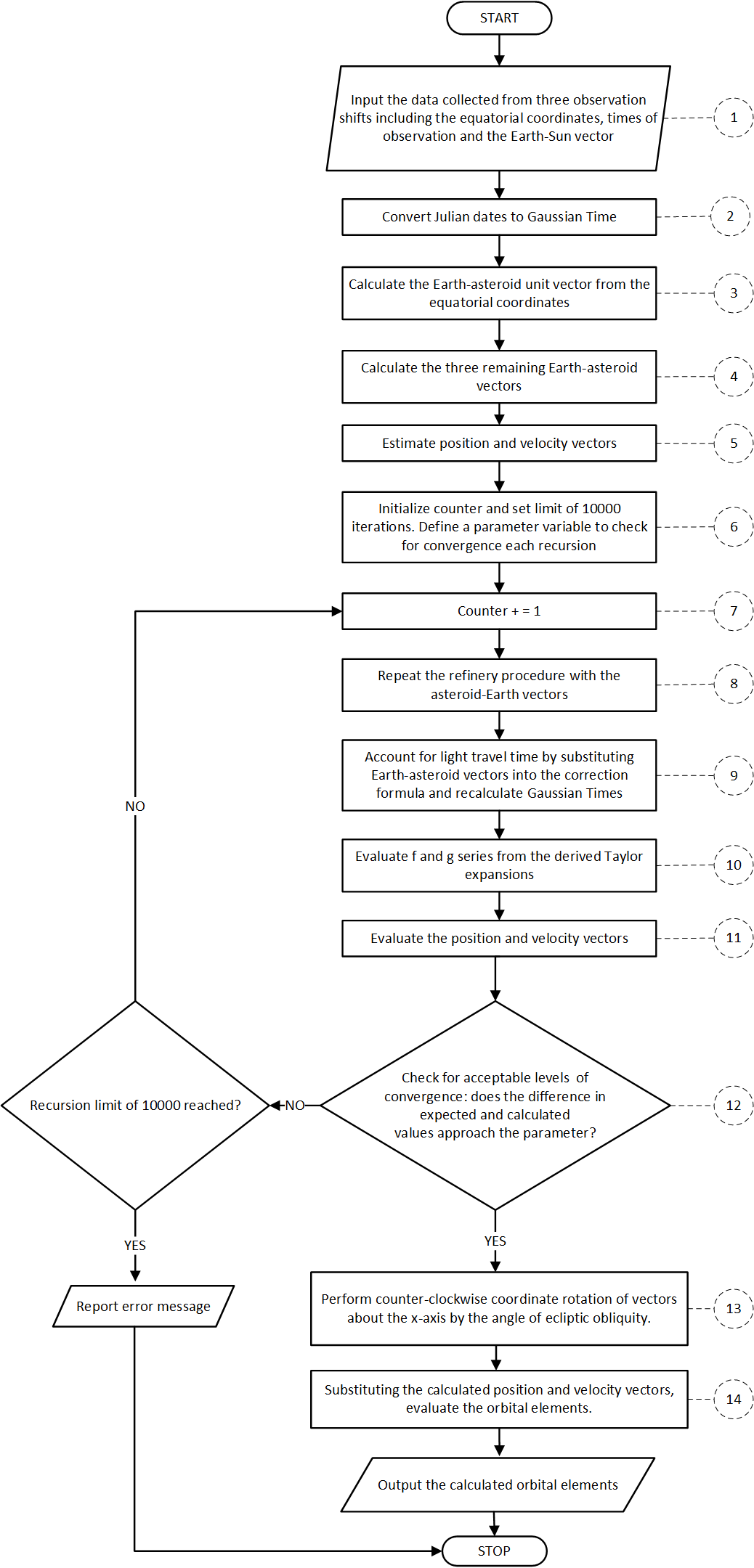}
\caption{
\label{fig3}
Algorithmic implementation of Gauss's Method summarized.}
\end{center}
\end{figure}
\FloatBarrier

\subsection{Preliminary Orbit Estimation}

For the initial computation of the position and velocity vectors for the second inspection, Gauss's method assumes a recursive approach analogously to the Newton-Raphson or Euler's Method. 

The initially required inputs are all obtained from a minimum of three observation shifts roughly equally spaced out in time intervals: the right ascensions ($\alpha$) and declinations ($\delta$) with the Julian dates for the first and the third observations are required for preliminary orbit determination as portrayed in Fig. 1 and Fig. 2. The intermediate results are iterated through each sequence of the program insofar the intermediaries exhibit convergence toward the expected value with minimal difference owing to the continuous increase in accuracy of Taylor series terms defining the position and velocity vectors. To break the loop, a parameter was defined accounting for when the difference decremented below a certain value (i.e. in this case $10^{-10}$). After the ideal convergence of the position and velocity vectors was proven, the final results given by the last recursion of the algorithm were outputted with the percent errors in the calculated values and the predicted values also outlined into a table. Fig. 3 represents the algorithmic development of the preliminary orbit determination implemented by programmed code. The detailed explanations relating each step are also given [Danby 1992]:

\FloatBarrier
\begin{enumerate}
  \item Specifically speaking, the input variables are the right ascensions, declinations, Julian dates, and Earth-Sun Vectors.
  \item The Universal Gravitational Constant (G) is only known to four places of precision. To avoid propagation of uncertainty, the conversion to Gaussian time is accomplished by the formula: 
$\tau$=kt, where t is the Julian date and $k=GM_{\odot}$, is a constant equal to 0.01720209847 $\frac{AU^{3}}{day^{2}}$. Such conversion derives more accurate results as k is well known to a higher degree of precision than the Gravitational Constant, thus eliminating the emergence of erroneous calculations over long iterations.
  \item For computation of the Earth-asteroid unit vector, the subsequent equation was applied from substitution of right ascensions and declinations pair for each astrometric extraction:
  \begin{multline}
      \hat{\rho}_{i} = \cos(\alpha_{i})\cos(\delta_{i})\hat{i}\\ + \sin(\alpha_{i})\cos(\delta_{i})\hat{j} + \sin(\delta_{i})\hat{k}; i = 1,2,3
  \end{multline}
  \item For the determination of the $\rho$ vectors, a number of variables must be defined first and evaluated using the $\hat{\rho}_{i}$ vectors:
  \begin{equation}
      \tau_{0}=k(t_{3}-t_{1}), \tau_{1}=k(t_{1}-t_{2}), \tau_{3}=k(t_{3}-t_{2})
  \end{equation}
  \begin{equation}
      a_{1} \approx \frac{\tau_{3}}{\tau_{0}}, a_{2} = -1, a_{3} \approx -\frac{\tau_{1}}{\tau_{0}}
  \end{equation}
  \begin{equation}
      D_{0} = \hat{\rho}_{1}\cdot (\hat{\rho}_{2}\times\hat{\rho}_{3})
  \end{equation}
  \[   \left\{
\begin{array}{ll}
      D_{1j} = (\vec{R_{j}}\times\hat{\rho_{2}})\cdot\hat{\rho_{3}}   \\
      D_{2j} = (\hat{\rho_{1}}\times\vec{R_{j}})\cdot\hat{\rho_{3}} & j = 1, 2, 3\\
      D_{3j} = \hat{\rho_{1}}\cdot (\hat{\rho_{2}}\times\vec{R_{j}}) \\
\end{array} 
\right. \]

Finally, the $\rho_{i}$ vectors can be found by:
\begin{equation}
    \rho_{i} = \frac{a_1D_{i1}+ a_2D_{i2} + a_3D_{i3}}{a_{i}D_{0}}; i = 1, 2, 3
\end{equation}
  \item The position and velocity vectors can be approximated via:
  \begin{equation}
      \vec{r_{i}} = \rho_{i}\hat{\rho_{i}}-\vec{R_i}
  \end{equation}
and
  \begin{equation}
      \dot{\vec{r_2}} \approx \frac{\vec{r_3}-\vec{r_1}}{\tau_0}
  \end{equation}
respectively.

  \item A recursive algorithm was defined to confirm the values converged toward the solution as a preset constant to the tenth decimal place; once the difference in the calculated value and the expected value submerged below the parameter, convergence was verified to have been achieved to the desired accuracy.
  
  \item Another parameter was defined to ensure the number of iterations did not exceed the limit.
  
  \item Using the recalculated values from each loop, refine the $\rho_{i}$ vectors using equation (5).
  
  \item Given the distance between the observer and the asteroid is non-negligible, the calculated Gaussian times needed to be accounted for the light travel time from the asteroid to the Earth rather than the apparent measurements. For this purpose, the light-correction formula was employed assuming the speed of light equalled 173.1446 $\frac{AU}{day}$ for i = 1,2,3:
  \begin{equation}
      t_i \Rightarrow t_{corrected} = t_{apparent} + \frac{\rho_i}{c}
  \end{equation}
  For the next stage, substitute the time corrected values into (2).
  
  \item Recompute the position and velocity vectors using (7) but with the time corrected $\rho_{i}$ vectors. In succession, substitute the f and g Taylor expansion series with interpolate values of the position and velocity vectors from each iteration. Note that the expressions are derived up to the fourth order and and are evaluated for i = 1, 3:
  \begin{multline}
      f_i = 1 - \frac{\tau_i^{2}}{24r^{3}} + \frac{\left(\vec{r}\cdot\dot{\vec{r}}\right)\tau_i^{3}}{2r^{5}}\\
      + \frac{\tau_i^{4}}{24r^{3}}\left[3\left(\frac{\dot{\vec{r}}\cdot\dot{\vec{r}}}{r^{2}}-\frac{1}{r^{3}}\right)-15\left(\frac{\vec{r}\cdot \dot{\vec{r}}}{r^{2}}\right)^{2}+\frac{1}{r^{3}}\right]
  \end{multline}
  \begin{equation}
      g_i = \tau_i - \frac{\tau_i^{3}}{6r^{3}} + \frac{(\vec{r}\cdot \dot{\vec{r}})\tau_i^{4}}{4r^{5}}
  \end{equation}
  
  \item Substitute the evaluated solutions into the equations relating the scalar multipliers of the vector components with the f and g series:
  \begin{equation}
      a_1 = \frac{g_3}{f_1g_3-f_3g_1}  
  \end{equation}
  \begin{equation}
    a_3 = \frac{g_1}{f_3g_1-f_1g_3}
  \end{equation}
  \begin{equation}
      b_1 = \frac{f_3}{f_3g_1-f_1g_3}
  \end{equation}
  \begin{equation}
      b_3 = \frac{f_1}{f_1g_3-f_3g_1}
  \end{equation}
 The position and velocity vectors can be calculated with the refined values:
  \begin{equation}
      \vec{r_2} = a_1\vec{r_1}+a_3\vec{r_3}
  \end{equation}
  \begin{equation}
      \dot{\vec{r_2}} = b_1\vec{r_1}+b_3\vec{r_3}
  \end{equation}
  
  \item If the difference between the refined vectors and predicted ones do not approach a minuscule amount, the iteration is ran again with the counter being incremented by one. Otherwise, the program exits the iteration once convergence is not reached beyond the delimited parameter.
  
  \item The calculated vectors are not in the correct coordinate system for orbital element determination. The algorithm must account for the tilt of the Earth by performing a transform of coordinates from the equatorial to the ecliptic plane. Firstly the obliquity must be found with the year t given to account for the dynamic tilt of the Earth. Having found $\varepsilon$, the coordinate transform can be applied:
 \begin{equation}
 \begin{pmatrix} 
    x' \\
    y' \\
    z'
  \end{pmatrix}
=
\begin{pmatrix}
  1 & 0 & 0 \\
  0 & \cos(\varepsilon) & \sin(\varepsilon) \\
  0 & -\sin(\varepsilon) & \cos(\varepsilon)
\end{pmatrix}
\begin{pmatrix}
  x \\
  y \\
  z
\end{pmatrix}
\end{equation}

  \item The transformed vectors are then substituted into the orbital element equations discussed in the proceeding sections.

\end{enumerate}

\subsection{Keplerian Orbital Elements}    
The orbital elements are comprised of six parameters as portrayed in Fig. 4 (el-Showk 2017), given by:
\begin{equation}
    v^{2} = \dot{\vec{r}}\cdot \dot{\vec{r}}
\end{equation}
\begin{equation}
    \vec{h} = \vec{r} \times \dot{\vec{r}}
\end{equation}
\begin{itemize}

  \item \textbf{Semi-major axis (a)}- If the orbit is elliptical (0$<$e$<$1) meaning the body is in a bound orbit, it can be modelled by the ellipse equation with the sun at one of the foci. It is a measure of the size of the orbit and is normally stated in AU, calculated by the succeeding equation:
  \begin{equation}
      \frac{x^{2}}{a^{2}}+\frac{y^{2}}{b^{2}}=1
  \end{equation}
  From such, the semi-major axis can be derived through:
  \begin{equation}
      a = \frac{1}{\frac{2}{r}-\frac{v^2}{\mu}}
  \end{equation}
 
  \item \textbf{Eccentricity (e)}- Asteroid orbits are basically elliptical (e $>$ 1), parabolic (e = 1), or hyperbolic (e $<$ 1). Simply put, the eccentricity is a measure of how elongated the orbit is, determining the shape of the orbit in its plane. In fact, eccentricity is dimensionless quantity and is calculated by the following formula, where the magnitude of the h-vector is substituted:
  \begin{equation}
      e = \sqrt{1-\left(\frac{h^2}{a}\right)}
  \end{equation}
The following consecutive three elements are angles describing the relative relationship between the body's orbital plane and the ecliptic plane of the Earth and the Sun.

  \item \textbf{Angle of Inclination (i)}- The angle defining the asteroid's orbital plane and the earth's ecliptic where 0\textdegree$<$i$<$180\textdegree, given by the formula:
  \begin{equation}
      i = \arccos\left(\frac{h_{z}}{h}\right)
  \end{equation}

  \item \textbf{Longitude of the Ascending Node ($\Omega$)}- The asteroid's inclined orbital plane intersects the ecliptic at the descending and the ascending nodes, separated by 180\textdegree. The right ascension, or the longitude, of the ascending node is the angle, relative to the ecliptic plane, among the Vernal Equinox and the ascending node measured eastwards over the x-axis. 
  \begin{equation}
      \cos(\Omega) = -\left(\frac{h_{y}}{h\sin(i)}\right)
  \end{equation}
  \begin{equation}
      \sin(\Omega) = \left(\frac{h_{x}}{h\sin(i)}\right)
  \end{equation}

  \item \textbf{Argument of Perihelion ($\omega$)}- Defined as the angle between the eccentricity and the node vectors, the argument of perihelion, or periapsis, is superimposed over the orbital plane, revealing how distant the line of ascending node is from the perihelion where 0\textdegree$<$$\omega$$<$360\textdegree. For its calculation, the true anomaly ($\nu$) must be found first: 
  \begin{equation}
      \cos(U) = \frac{x\cos(\Omega)+y\sin(\Omega)}{r}
  \end{equation}
  \begin{equation}
      \sin(U) = \frac{z}{r\sin(i)}
  \end{equation}
  \begin{equation}
      \cos(\nu) = \frac{1}{e}\left(\frac{a(1-e^2)}{r}-1\right)
  \end{equation}
  \begin{equation}
      \sin(\nu) = \frac{1}{e}\left(\frac{a(1-e^2)}{h}\cdot\frac{\vec{r}\cdot \dot{\vec{r}}}{r}\right)
  \end{equation}
  \begin{equation}
      \omega = U - \nu
  \end{equation}

  \item \textbf{Mean Anomaly (M)}- Assuming a circular orbit, the factor provides a predicted position as an angle in degrees measured from the perihelion as the baseline, enabling the position of the satellite to be calculated at any given time. To find the mean anomaly, the eccentric anomaly (E) must be found first by the following: 
  \begin{equation}
      E = \arccos\left[\frac{1}{e}\left(1-\frac{r}{a}\right)\right],
  \end{equation}
 note that E is calculated within the same quadrant as $\nu$. Now the mean anomaly can be found as the Kepler's Equation:
  \begin{equation}
      M = E - e\sin(E)
  \end{equation} 

  \begin{figure}[H]
  \begin{center}
  \includegraphics[width=14.5pc]{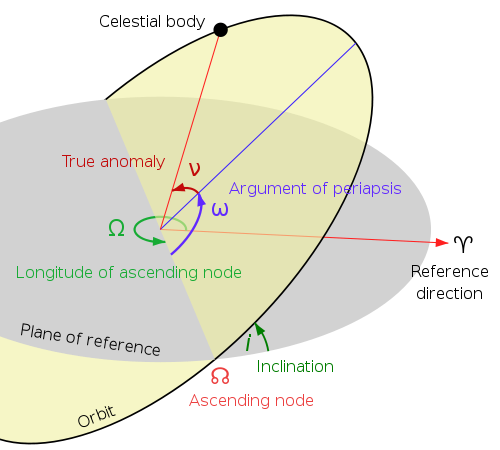}
  \caption{\centering
  \label{fig1}
  Illustration of Keplerian orbital elements (Courtesy of: https://en.wikipedia.org/wiki/Orbital\_elements\#/media/File:Orbit1.svg).}
  \end{center}
  \end{figure}
\end{itemize}

\subsection{Overall Orbit Determination}
Substituting the vectors calculated in step one, now the orbital elements can be calculated for the second observation date. As encapsulated by the following outline of the algorithm, the final procedure utilizes a variety of widely accepted techniques and concepts including the orbital-energy-invariance law, conservation of mechanical energy and momentum in order to solve for the orbital elements.

Using Gauss's Method, programs were compiled in Python (Python Software Foundation, 2020) to determine the orbit of our asteroid in code form. We implemented object-oriented programming to ensure the re-usability of orbit determination (Cielaszyk \& Wie 1996) programs in later test cases. The coding procedure was divided across multiple Orbit Determination (OD) programs:

\begin{itemize}
  \item \textbf{OD1}- This code was inputted with the right ascensions and declinations alongside the position and velocity vectors to calculate the orbital elements. 
  \item \textbf{OD2}- An iterative solver later incorporated into OD4. It received as inputs the velocities of two test particles and outputted the time taken for one particle to catch up to the other in a while loop.
  \item \textbf{OD3}- Received right ascension, declination and the Earth-Sun vectors as inputs.  Using these values, we evaluated the Earth-Asteroid vector ($\rho$) and its unit vector ($\hat{\rho}$). Appropriate coordinate conversion of right ascension and declination to Cartesian was also taken into consideration, and hence applied.
  \item \textbf{OD4}- Utilizing the coordinate conversion of OD3 and the iterative solver of OD2, a program was composed calculate the position ($\vec{r}$) and velocity vectors ($\dot\vec{r}$) using Gauss's method.
  \item \textbf{OD5}- Taking the orbital elements alongside the position and velocity vectors, the algorithm generated an ephemeris as an output for a specified time.
  \item \textbf{OD6}- Combined OD4 and OD1 in order to have one program package that takes the right ascension and declination to extract the orbital elements: in other words the complete orbit determination implementation in an algorithmic format.
\end{itemize}

\section{Experimental Test Case-Study}
    For this project, the orbit of 12538 (1998 OH) was analyzed (Asteroid  (12538)  1998  OH,  2020). The said asteroid is located within Zone I of the Asteroid Belt (a$<$2.5), and classified as an Apollo asteroid with a semi-major axis greater than 1 AU and perihelion distances less than the Earth's aphelion distance (Q$<$1.017 AU). Gauss's Method was executed, which required three position vectors roughly equally spaced out in time, details of which are outlined and extended appropriately in the orbit determination section.

\subsection{Observations and Image Processing}
\subsubsection{Data Acquisition}
Our observational data was collected at the Sommers Bausch Observatory at the University of Colorado Boulder operating the PlaneWave CDK telescope set-up: specifically, the type Corrected Dull Kirkham with a 20 inch mirror. These telescopic instruments were fitted with a SBIG STF-8300m (KAF 8300 Sensor) kodex CCD camera and the host computers used to maneuver the telescopes utilized TheSkyX (https://theskyx-professional-edition.software.informer.com/10.1/) software to capture and store the images. We had multiple underlying observations throughout the six-week duration, some having been undertaken in more favorable conditions than others, the succeeding Table 1 summarizes our observation record.

\begin{table*}[t]
\tabularfont
\centering
\setlength{\tabcolsep}{25pt}
\caption{\centering{Asteroid's Image observations using the telescope.}}
\vspace{3mm}
\begin{tabular}{c|c|c|c}\label{Observations}
Date & Time (MDT) & Number of images taken & Quality of Images \\
\hline
6/26/19 & 22:45-00:00 & 5 darks, 10 lights & Clear, Asteroid Identified \\
6/28/19 & 22:45-00:00 & 5 darks, 10 lights & Clear, Asteroid Identified \\
7/2/19 & 00:00-01:15  & 5 darks, 10 lights & Clear, Asteroid Identified \\
7/6/19  & 22:45-00:00 & 5 darks, 10 lights & Clear, Asteroid Identified \\
7/15/19 & 21:00-22:45 & 5 darks, 10 lights & Clear, Asteroid Identified \\
7/18/19 & 21:00-22:45 & 5 darks, 10 lights & Clear, Asteroid Identified\\
\hline
\end{tabular}
\end{table*}

\begin{table*}
\tabularfont
\setlength{\tabcolsep}{15pt}
\begin{center}
\caption{\centering Photometry performed on Asteroid 1998 OH across the four instances of observation spans.}
\vspace{3mm}

    \begin{tabular}{c|c|c|c}\label{Photometry}
    Date MM/DD/YY (MDT) & Expected Value & Calculated Value (2 d.p.) & Deviation from Expected \\
    \hline
    06/26/19 & 16.49 & 16.34 & -0.15 \\
    07/06/19 & 16.88 & 16.41 & -0.47 \\
    07/15/19 & 17.20 & 17.07 & -0.13 \\
    07/18/19 & 17.22 & 17.50 & +0.28 \\
    
    \hline
    \end{tabular}
    \end{center}
\end{table*}

\subsubsection{Image Reduction}
For the astrometry reduction process (Urban \& Seidelmann 1992), the AstroImageJ software (https://www.astro.louisville
.edu/software/astroimagej/) was used. Feeding 10 lights, 5 flats, and 5 dark images from each observation shift into the application, it returned the reduced images which were used afterwards to locate 1998 OH. The reduction process required the starfield images to be first corrected using a flat, which were normalized, and the background images to be subtracted. To these ends, the software automatically applied the following formula:
\begin{equation}
\mathrm{reduction} =\frac{\mathrm{light}-\mathrm{dark}}{\mathrm{flat}}
\end{equation}
to each pixel count. In conjunction with AstroImageJ, the software application SAOImage DS9 (http://ds9.si.edu/site/Home.html) was operated to conduct the selection of reference stars. Excluding reference stars with magnitude less than 14 and cross-collating the two image plates from DS9 and the reduced images, at least ten reference stars were located and selected to run the built-in plate-solve function, for which to run effectively required the stars to be sufficiently distanced and luminous insofar as to negate over-saturation; in this case, the maximum permissible pixel count was delimited to 45000 counts due to the extremity of the CCD camera's dynamic range employed in the telescope apparatus. Subsequent of the reference star selection, the plate-solve function was executed, returning the right ascensions and declinations alongside their standard Cartesian coordinates of every object in the starfield including the identified asteroid. The primal purpose of the plate-solving was to derive the centroid of the asteroid, which hence enabled us to perform astrometry on all the process reduced images. Fig. 5 summarizes the final step in the astrometry procedure when the asteroid has been identified after the plate-solving, labeled in red. 

\begin{figure}[h]
\begin{center}
\includegraphics[width=19.5pc]{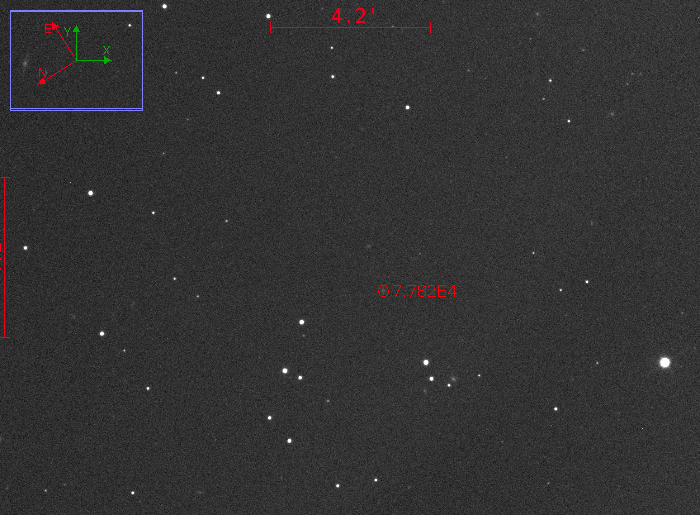}
\caption{\centering
1998 OH located within red circle.}
\end{center}
\end{figure}

    In order to decipher the magnitude of the asteroid in images, photometry was executed . The magnitude of the asteroid is given by the following formula:
    \begin{equation}
     m_{ast} = m_{star} - 2.5\log\left(\frac{F_{ast}}{F_{star}}\right)
    \end{equation}
    AstroImageJ returned the right ascensions and declinations of reference stars that were compared to stars with known magnitudes in the DS9 optical UCAC4 catalog, from which the flux values and the apparent magnitudes were substituted into the photometry equality alongside the asteroid's flux obtained from JPL (Jet Propulsion Laboratory) Horizons (https://ssd.jpl.nasa.gov/?horizons=), consequently enabling the calculation of 1998 OH's apparent magnitude. This process was repeated for four observations, the results of which are given in Table 2.  Note a maximum of +/-0.5 deviation from the expected value given by JPL Horizons was permitted and acquired results fit the permitted range as shown. Evidenced by the trend in the deviations, the photometric results imply some sort of a light curve pattern. Due to the limited time period over which the photometry was conducted, it is arduous to investigate whether the cause was a simple coincidence or otherwise. The next step was to estimate the size of 1998 OH, for which the following formula (Mark 2016), where D is the diameter of the asteroid in kilometers, $p_v$ is the geometric albedo, and H is the absolute magnitude, was appropriately applied. Both the geometric albedo and the absolute magnitude were obtained from JPL Horizons and are unitless:
\begin{equation}
    D(p_v, H) = \frac{1329}{\sqrt{p_v}}\cdot 10^{-0.2H}
\end{equation}
Using JPL Horizons, the values were found to be 0.232 and 15.8 for the albedo and absolute magnitude respectively (note that both values are dimensionless). Evaluating D for these two values, the diameter of the asteroid gauged to be 1.91 km (2 d.p.), which is in agreement with previously published data pertaining to this asteroid, estimating the size to be within the range of 1.66+/-0.329 km (Asteroid (12538) 1998 OH, 2020).

\subsubsection{Evaluation of the reliability of astrometric applications}
\label{sec1}

Lately, the orbit computations have been spurred by the advent of state-of-art statistical schemes rather than the deterministic pursuits (Bowell {\em et al.} 2002) as far as the small-scale problems are concerned. Due to the inherent, erroneous nature of astrometry performed by AstroImageJ, it was imperative to run a manual astrometry program on Python (Python Software Foundation 2020) called the Least Squares Plate Reduction (LSPR) inputted with the Cartesian coordinates taken from AstroImageJ, whereas right ascensions and declinations were retrieved from DS9 (the algorithmic approach implemented in the Python programs are reported in the Appendix A). By running the same procedure individually for thirty reference stars computed from the USNO-CAC4 image server catalog, and feeding the data into LSPR python program, it became feasible to evaluate the right ascension and declination of the asteroid, hence utilize those measurements to calculate uncertainties and standard deviations of the values to reach a conclusion whether the embedded plate-solving function of the AstroImageJ software was truly accurate or not. For such error determination, the sample images captured between 4:00 and 7:00 Coordinated Universal Time (UTC) on June 26, 2019 were availed. The total number of reduced images compiled into a chronological image sequence amounted to ten plates for extensive error analysis using thirty reference stars loaded into AstroImageJ.

Subsequent the retrieval of a DS9 USNO-UCAC4 catalog image (centered about the coordinates: right ascension = 254.40$^{\circ}$, declination = 35.84$^{\circ}$) with area of the plate constituting twenty by twenty arc-minutes, twenty-eight out of thirty reference stars were chosen for comparison with the plates on AstroImageJ.

Recording the right ascensions and declinations of the stars from both the DS9 image server and the plate-solved images of the AstroImageJ, residuals were calculated by the following formula, enabling the enumeration of the standard deviation and the mean via the predicted value minus the actual value. Table 4 summarizes the test residuals, alongside the calculated average and the uncertainty, succinctly encapsulated by Fig. 6 that follows. For executing comparison tests, we imported the right ascensions and declinations into Google Sheets (https://www.google.com/sheets/about/).

We first calculated residuals, following the equation (predicted-known):
\begin{equation}
    \sigma_{\alpha} = \alpha_{ASJ}-\alpha_{DS9}
\end{equation}
\begin{equation}
    \sigma_{\delta} = \delta_{ASJ}-\delta_{DS9}
\end{equation}

\noindent then averaged the residuals and calculated the standard deviation of the residuals alongside the Standard Error of the mean:
\begin{equation}
    SE = \frac{\sigma}{\sqrt{N}}
\end{equation}

\begin{figure*}[t]
\centering
\begin{center}
\includegraphics[width=20pc]{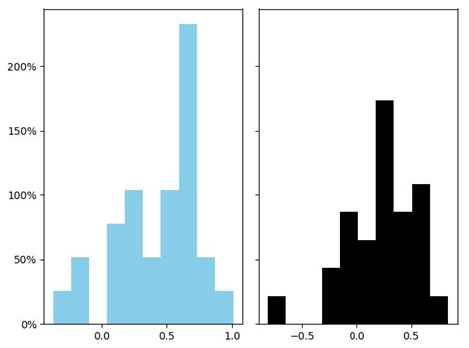}
\caption{\centering
\label{fig5}
Right Ascension and declination residuals shown in blue and black respectively.}
\end{center}
\end{figure*}

It must be noted the right ascensions and declinations are reported in decimal degrees whereas the residuals are displayed in arcseconds with a precision of two decimal places as reported in Table 4. As portrayed, the mean residuals are within the standard deviation for both $\alpha$ and $\delta$. Figure \ref{fig5} shows the distribution of the residuals of $\alpha$ and $\delta$, respectively.

\section{Results, Analysis, and Evaluation}
To verify the accuracy of our OD code, test case values from JPL Horizons were taken as benchmarking matrices to enable relative comparison. Theoretically speaking, the orbit determination conducted by the programs should be more representative than the predicted values provided by JPL Horizons; henceforth, this comparison was intended to be more of a sanity check. Following the rigorous error analysis and substantiation of the reliability regarding the right ascension and declination taken from AstroImageJ, orbital elements could be measured. Initially, the obtained results via astrometry for three of the earlier observations were chosen as the initial, middle, and final values for OD4. These values are shown in Table 3, with the components of $\vec{R}$ in Table 5 displaying the heliocentric coordinates of Earth.

\subsection{Monte Carlo Histograms}
For the sake of substantiating the final results obtained pertaining to the orbital elements, Monte Carlo sampling (Li M {\em et al.} 2018; Song J. \& Xu G. 2017) was employed to further affirm the calculated values were within an acceptable variance derived from the mean equal to averaged observational coordinates, and the standard deviation developed by Least Squares Plate Reduction in Table 4. The test case was implemented for the observation date 2019-Jul-04 05:12:26.64 UTC to devise the standard deviations in individual orbital elements. Sequentially, 10,000 independent samples of the six astrometric $\left(\alpha,\delta \right)$ coordinates was constructed by a randomized selection of values from a normal distribution of inputs, for each of which the orbital elements were discretely defined and centered at the astrometric coordinates. As error analysis had only been done for the middle observation, same error bars for each observation were applied. The produced orbital element data was arranged and composed into histograms with 25 evenly spaced bins spanning the range of the values calculated with the bin frequencies normalized, urging the distributions in test outputs to emerge adhering to the standard deviation. The calculated central orbital elements obtained from OD5 for 1998 OH at 2019-Jul-04 05:12:26.64 are quoted in Table 6 alongside their standard deviations obtained from the Monte Carlo histograms represented by Figs. 7, 8, and 9.

\begin{table}[t]
\centering
\tabularfont
\scriptsize
\caption{RA and DEC from ASJ astrometry.}

\begin{tabular}[t]{c|c|c}
\hline
Time (UTC) & $\alpha$ (HH:MM:SS.SS) & $\delta$ (DD:MM:SS.SSSS)  \\ %
\hline
2019-Jun-27 05:27:36.35 & 15:01:46.8700 & 35:04:02.60 \\
2019-Jul-04 05:12:26.64 & 15:22:14.7864 & 32:36:35.01 \\
2019-Jul-10 07:14:35.69 & 15:37:29.0028 & 30:26:26.07 \\
\hline
\end{tabular}
\end{table}

\FloatBarrier
\begin{table*}[t]
\centering
\tabularfont
\caption{\centering Right ascensions, declinations, residuals, and mean standard deviation of 1998 OH obtained from AstroImageJ and DS9.}
\begin{tabular}[t]{cccc|cc} \label{asdasd}
$\alpha$ DS9 $(^{\circ})$ & $\delta$ DS9 $(^{\circ})$ & $\alpha$ Plate & $\delta$ Plate & $\alpha$ Residuals (")& $\delta$ Residuals (") \\
\hline
225.41357360 & 35.07893530 & 225.4136700 & 35.0789310 & 3.47E-01  & -1.55E-02 \\
225.43261620 & 35.06471170 & 225.4327950 & 35.0646710 & 6.44E-01  & -1.47E-01 \\
225.42663420 & 35.03733870 & 225.4269150 & 35.0374250 & 1.01E+00  & 3.11E-01  \\
225.39209030 & 35.03018340 & 225.3922800 & 35.0301930 & 6.83E-01  & 3.46E-02  \\
225.39930360 & 35.01246280 & 225.3995400 & 35.0124530 & 8.51E-01  & -3.53E-02 \\
225.34375000 & 35.04067200 & 225.3438150 & 35.0406940 & 2.34E-01  & 7.92E-02  \\
225.33913590 & 35.08566450 & 225.3393450 & 35.0857530 & 7.53E-01  & 3.19E-01  \\
225.35213680 & 35.11589500 & 225.3522900 & 35.1158370 & 5.52E-01  & -2.09E-01 \\
225.50857620 & 35.14403980 & 225.5087700 & 35.1441080 & 6.98E-01  & 2.46E-01  \\
225.47262390 & 35.09510030 & 225.4726650 & 35.0951780 & 1.48E-01  & 2.80E-01  \\
225.50594740 & 35.09469980 & 225.5060250 & 35.0948460 & 2.79E-01  & 5.26E-01  \\
225.49386420 & 35.11576780 & 225.4939350 & 35.1157960 & 2.55E-01  & 1.02E-01  \\
225.49091920 & 35.12008140 & 225.4910850 & 35.1201590 & 5.97E-01  & 2.79E-01  \\
225.44081000 & 35.14755120 & 225.4409850 & 35.1476770 & 6.30E-01  & 4.53E-01  \\
225.49820090 & 34.97787450 & 225.4982250 & 34.9778210 & 8.68E-02  & -1.93E-01 \\
225.56154420 & 34.98546640 & 225.5617050 & 34.9855570 & -5.79E-01  & 3.26E-01  \\
225.52300530 & 34.93384140 & 225.5229750 & 34.9339780 & -1.09E-01 & 4.92E-01  \\
225.40684060 & 34.94401730 & 225.4070250 & 34.9442010 & 6.64E-01  & 6.61E-01  \\
225.38984330 & 34.90627640 & 225.3900000 & 34.9065090 & 5.64E-01  & 8.37E-01  \\
225.34129330 & 35.12058640 & 225.3413700 & 35.1206950 & 2.76E-01  & 3.91E-01  \\
225.59288620 & 34.99222810 & 225.5930850 & 34.9923140 & 7.16E-01  & 3.09E-01  \\
225.32551920 & 35.15187870 & 225.3256800 & 35.1520100 & 5.79E-01  & 4.73E-01  \\
225.51795030 & 34.99899340 & 225.5180550 & 34.9989550 & 3.77E-01  & -1.38E-01 \\
225.58675500 & 34.98819780 & 225.5869350 & 34.9883670 & 6.48E-01  & 6.09E-01  \\
225.38498710 & 35.09970560 & 225.3850350 & 35.0998580 & 1.72E-01  & 5.49E-01  \\
225.34419150 & 35.05953230 & 225.3441600 & 35.0596010 & -1.13E-01 & 2.47E-01  \\
225.32912450 & 35.01796950 & 225.3293100 & 35.0181330 & 6.68E-01  & 5.89E-01  \\
225.34703800 & 34.92976230 & 225.3469350 & 34.9295370 & -3.71E-01 & -8.11E-01 \\
\hline
Mean & - & - & - & 3.61E-01 & 2.344E-01 \\
Standard Deviation & - & - & - & 4.15E-01 & 3.440E-01 \\
\hline
\end{tabular}
\end{table*}

\FloatBarrier


\begin{table*}[t]
\centering
\tabularfont
\caption{\centering $\vec{R}$: Earth Sun Vector obtained from AstroImageJ astrometry.}
\vspace{3mm}
\begin{tabular}{c||c|c|c} 
\hline

Time (UTC) & $\vec{x}$ (au) & $\vec{y}$ (au) & $\vec{z}$ (au)    \\
\hline
2019-Jun-27 05:27:36.35 & -0.089389357115396 & 0.929122994534851 &  0.40273004198922 \\
2019-Jul-04 05:12:26.64 & -0.206375720170234 & 0.913481492972422 & 0.395953433102251 \\
2019-Jul-10 07:14:35.69 & -0.305959631358509 & 0.889587295448561 &  0.385598325169787 \\
\hline
\end{tabular}
\end{table*}

\begin{table*}[t] \label{calcorbel}
\tabularfont
\centering
\caption{\centering Comparison of orbital elements calculated from our designed program and from JPL Horizons.}
\vspace{3mm}
\begin{tabular}{l|ccccc}
\hline
Parameter Name & JPL & Mean Calculated & & Uncertainty & Percent Discrepancy (\%) \\
\hline
Semi-major Axis (a) & 1.541852 AU & 1.51358 AU & $\pm $  & 0.037 AU & 1.833163 \\
Eccentricity (e) & 0.406025 & 0.396045 & $\pm$ & 0.013 & 2.457992 \\
Inclination (i) & 24.526318$^{\circ}$ & 24.288149$^{\circ}$ & $\pm$  & 0.327$^{\circ}$ & 0.971078 \\
Longitude of Ascending Node ($\Omega$)& 220.744933$^{\circ}$ & 221.049404$^{\circ}$ & $\pm$ & 0.401$^{\circ}$ & 0.137929 \\
Argument of Perihelion ($\omega$)& 321.737397$^{\circ}$ & 320.920565$^{\circ}$ & $\pm$ & 1.10$^{\circ}$ & 0.253881 \\
Mean Anomaly (M)& 42.384887$^{\circ}$ &  43.883072$^{\circ}$ & $\pm$ & 2.05$^{\circ}$ & 3.534715  \\
\hline
\end{tabular}
\end{table*}

\begin{figure*}[t]
\centering
\centerline{
\includegraphics*[width=13pc]{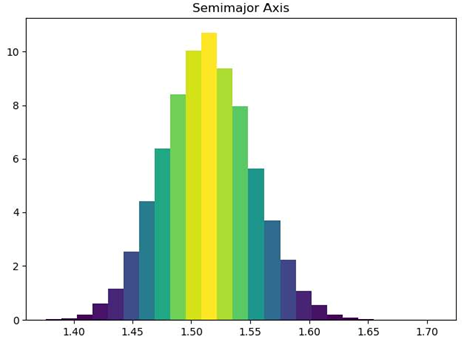}
\includegraphics*[width=13pc]{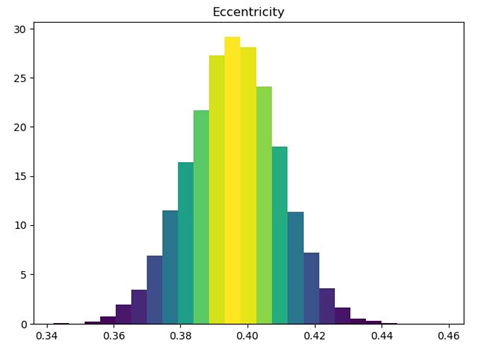}
}
\caption{\centering
Monte Carlo histograms showcasing semi-major axis \& eccentricity (the x-axis of the left graph has units of AU, and the right is unitless).}
\end{figure*}
\begin{figure*}[t] \label{inc,Omega}
\centering
\centerline{
\includegraphics[width=13pc]{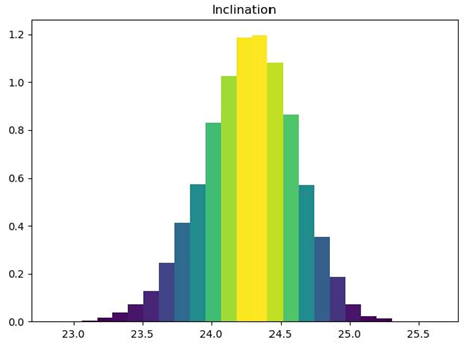}
\includegraphics[width=13pc]{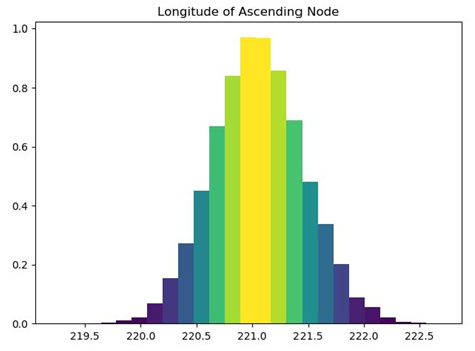}
}
\caption{\centering
Monte Carlo histograms showcasing inclination \& longitude of ascending node (the x-axes of both graphs have units of degrees).}
\end{figure*}
\begin{figure*}[t] \label{omega,M}
\centering
\centerline{
\includegraphics[width=13pc]{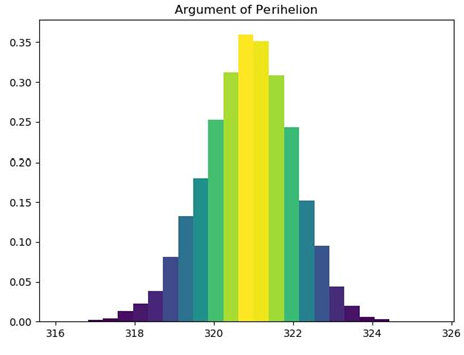}
\includegraphics[width=13pc]{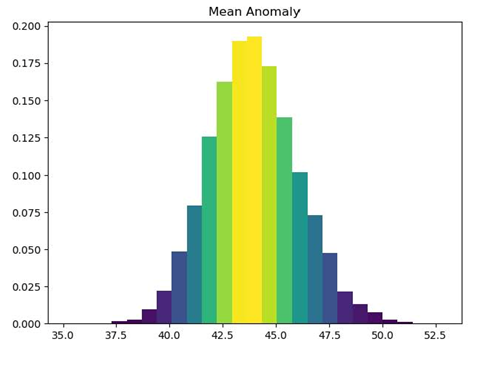}
}
\caption{\centering
\label{a}
Monte Carlo histograms showcasing the argument of perihelion \& mean anomaly (the x-axes of both graphs have units of degrees).}
\end{figure*}

\clearpage

\subsection{Ephemeris Generation}
Hitherto the prime focus of the study centered the AstroImageJ as its most potential causation of error in result generation notwithstanding the probable erroneous nature of the OD programs. For the purpose of ensuring self-consistency in manual orbital determination, the Gauss's Method was reverse-engineered to generate a ephemeris of celestial coordinates [Heafner 1999]. 

The reciprocal process of the orbit determination method can be outlined as follows, implemented in the said OD5 algorithm [Danby 1992]:

\begin{enumerate}
   \item Initialize the procedure by inputting the Julian date,  six orbital elements (a, e, i, $\Omega$, $\omega$, and M) and the components of the Earth-Sun vector. It must be noted however the mean anomaly alludes to a time-dependent quantity; thus in order to derive an accurate value for the variable, M must be rectified using the Julian date to calculate T, the reference epoch. Afterwards, the eccentric anomaly inversely derived from Kepler's Equation via the Newton-Raphson Method enables orbital coordinate conversion.

   \begin{itemize}
      \item For the said method, the refined version of mean anomaly was defined as:
     \begin{equation}
         M_{refined} = n(t-T)
     \end{equation}
     where n, the mean angular motion, is equal to
     \begin{equation}
         n = k\sqrt{\frac{\mu}{a^{3}}}
     \end{equation}
     
     and the constants k and $\mu$ were equated to 0.01720209894 and 1.00 respectively; and T could be found in units of days by the rearrangement of the Kepler's formula where t represents the date for which the ephemeris is desired, and t$_2$ the date of observation:
     \begin{equation}
         T = t_2 - \frac{M_{crude}}{nk}
     \end{equation}
     \item After the determination of M at ephemeris time, the accurate eccentric anomaly could be found with the iterative nature of Newton's Method. In general form:
     \begin{equation}
         E_{n+1} = E_{n} - \frac{f(E_{n})}{f'(E_{n})}
     \end{equation}
     Henceforth, the initial guess would just be equal to M. Sequentially, the function must be substituted as $M-(E_{n}-e\sin(E_{n}))$ and its derivative as $e\cos(E_{n})-1$ for an arbitrary number of times before its convergence to the solution:
     \begin{equation}
         E_{n+1} = E_{n}-\left(\frac{M-E_{n}-e\sin(E_{n})}{e\cos(E_{n})-1} \right)
     \end{equation}
   \end{itemize}
   \item The accurate enumeration of the eccentric anomaly enables for the Cartesian coordinates of the asteroid, with the x-axis coinciding along the perihelion, to be computed via:
   \begin{equation}
\begin{pmatrix}
    x \\
    y \\
    z
  \end{pmatrix}
=
\begin{pmatrix}
  a\cos{E}-ae \\
  a\sqrt{1-e^{2}}\sin{E} \\
  0
\end{pmatrix}
\end{equation}

    \item The Cartesian values must then be transformed to ecliptic ones by the coordinate rotation from the negative $\omega$, i, and $\Omega$ values about the z, x and z axes, respectively:
\begin{multline}
\begin{pmatrix}
    x_1 \\
    y_1 \\
    z_1
\end{pmatrix}
=
\begin{pmatrix}
  \cos{\Omega} & -\sin{\Omega} & 0 \\
  \sin{\Omega} & \cos{\Omega} & 0 \\
  0 & 0 & 1
\end{pmatrix}
\begin{pmatrix}
  0 & 0 & 0 \\
  0 & \cos{i} & -\sin{i} \\
  1 & \sin{i} & \cos{i} 
\end{pmatrix}
\\
\begin{pmatrix}
  \cos{\omega} & -\sin{\omega} & 0 \\
  \sin{\omega} & \cos{\omega} & 0 \\
  0 & 0 & 1
\end{pmatrix}
\begin{pmatrix}
  x \\
  y \\
  z
\end{pmatrix}
\end{multline}

   \item Establish equatorial-ecliptic conversion via the rotational matrix and obliquity:
\begin{equation}
\begin{pmatrix}
    x_2 \\
    y_2 \\
    z_2
\end{pmatrix}
=
\begin{pmatrix}
  0 & 0 & 0 \\
  0 & \cos{\varepsilon} & -\sin{\varepsilon} \\
  1 & \sin{\varepsilon} & \cos{\varepsilon} 
\end{pmatrix}
\begin{pmatrix}
    x_1 \\
    y_1 \\
    z_1
\end{pmatrix}
\end{equation}

   \item Calculate the Earth-asteroid range vectors via simple vector addition: $\vec{\rho} = \vec{r} + \vec{R}$, and hence the rho-hat vectors from $\hat{\rho} = \frac{\vec{\rho}}{\rho}$.
   
   \item Lastly, substitute the range vectors for right ascension and declination into:
   \begin{equation}
       \sin{\delta} = \hat{\rho_z}
   \end{equation}
   \begin{equation}
       \cos{\alpha} = \frac{\hat{\rho_x}}{\cos{\delta}} \quad , \quad \sin{\alpha} = \frac{\hat{\rho_y}}{\cos{\delta}}
   \end{equation}
\end{enumerate}

\begin{table*}[t] \label{sanity}
\tabularfont
\centering
\caption{\centering Comparison of right ascension and declination values recorded from AstroImageJ and predicted ones.}
\vspace{3mm}
  \begin{tabular}{l|c|c|c|c}
\hline
Julian Time & Actual RA ($^{\circ}$) & Generated RA ($^{\circ}$) & RA Residual ($^{\circ}$) & Percent Error (\%)\\
\hline
2458671.708030 & 232.493955 & 232.5059964 & 0.0120414 & 0.00517921 \\
2458680.655547 & 237.67263 & 237.7050108 & 0.0323808 & 0.0136241 \\
2458683.644220 & 239.257905 & 239.2947145 & 0.0368095 & 0.0153849 \\
\hline
Julian Time & Observed Dec ($^{\circ}$) & Generated Dec ($^{\circ}$) & Dec Residual ($^{\circ}$) & Percent Error (\%)\\
\hline
2458671.708030 & 31.544784 & 31.54789028 & 0.0031063 & 0.0013361 \\
2458680.655547 & 28.356835 & 28.36485058 & 0.0080156 & 0.0033725 \\
2458683.644220 & 27.297882 & 27.30743264 & 0.0095506 & 0.0039918 \\
\hline
\end{tabular}
\end{table*}

Table 7 summarizes the results comparison between the derived calculation from the ephemeris generation algorithm and the astrometric predictions. As delineated, the percent errors are insignificant to the hundredth and thousandth places, confirming the reliability of our programmed OD5's potential in generating ephemerides. A plausible implication worth considering is the JPL's online database may not be up-to-date, which could have resulted in an obsolete ephemeris generation system; to mitigate the such, differential correction [Heafer 1999] could pay dividends to the resultant calculations by minimizing the difference in observed and calculated orbital elements, incrementing percent accuracy. For differential correction, a minimum of four adequate observations would be required with a total of eight positional data points, and repeating the ephemeris generation procedure afterwards. Using linear regression, a best fit can then be computed for the orbital parameters more accurately.

\subsection{Plot of Asteroid Orbit}

\begin{figure}[h]
\centering
\begin{center}
\includegraphics[width=21pc]{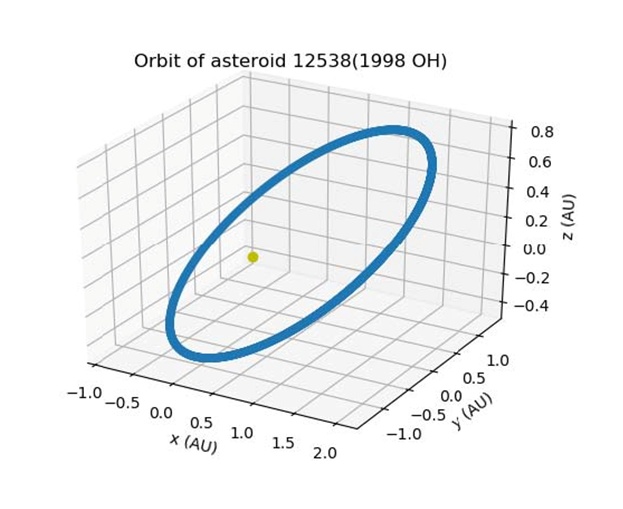}
\caption{\centering
Our Python program's generated 3D model of 1998 OH's orbit with the Sun in yellow.}
\end{center}
\end{figure}

\begin{figure}
\centering
\begin{center}
\includegraphics[width=18pc]{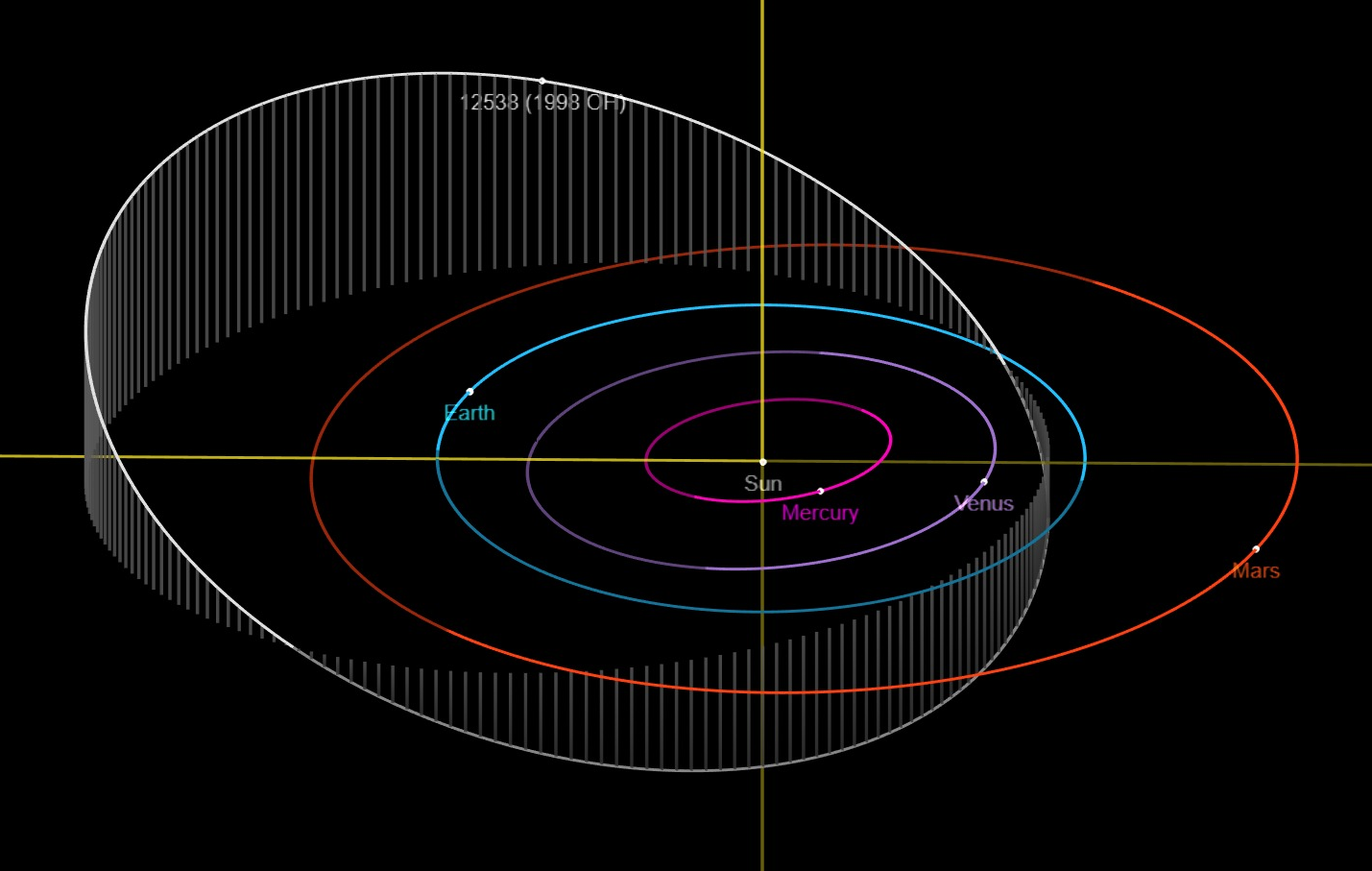}
\caption{\centering
 Courtesy of: NASA - Jet Propulsion Laboratory, California Institute of Technology, https://ssd.jpl.nasa.gov/sbdb.cgi?sstr=12538;old=0;orb=1;cov=0;
 log=0;cad=0\#orb}
\end{center}
\end{figure}

Finally, a program was composed to generate a 3-D orbit of our asteroid displayed by Fig. 10 from the calculated orbital elements in Table 6 via a convenient plotting method given in Appendix B. The rather large inclination and eccentricity of this orbit shown parallel to the xy plane matches our calculated predictions. Fig. 11 displays the orbital path of 1998 OH with respect to the relative trajectories of other planets.

Undoubtedly, the major cause of the complicating errors (Table 6) can be attributed to the asteroid's especially inclined orbit. Starting in mid-July, about halfway through the data collection, the asteroid began to rise above the ecliptic plane rapidly, something that the OD codes were not able to account for. This initially rendered an accurate extrapolation impossible, let alone reasonable, which propagated through to a further inability to calculate the orbital elements. Eventually, however, a set of vectors were obtained through combination of the observational data and the evidence of the asteroid's tendency to rise above the ecliptic is undeniably present in those observations, which would contribute to one main source of error for such results. As demonstrated in Table 6, the calculated and expected orbital elements are found to be comparable. The significant errors are in the semi-major axis (a), eccentricity (e), and mean anomaly (M) values. These three elements all depend on the value of the position vector, which exhibited a 3.2\% difference in the x component, which is likely the cause of this discrepancy. Theoretically, since the position vector is directly calculated from an observation as opposed to extrapolated by JPL, it should be more accurate. For the sake of a self consistency check, as displayed in Table 8, a very small error between the observed coordinates and those produced by the ephemeris generation program can be shown, indicating the calculated orbital elements were indeed accurate.

\subsection{Asteroid Simulations}

Notwithstanding the majority of orbit determinations being done through mathematical formulaic implementations in Python programming, a crux of the subject matter also entailed a numerical integration simulation to predict the future of the 1998 OH asteroid and its behavioral orbit dynamics; however, such task proved to be an arduous one owing to the concept of two-body approximations, which maintains itself at an optimal estimate assuming no presence of external perturbations. To model the test particles in a heliocentric orbit with multiple bodies, such an approximation would be expected to fall short of deriving any reliable, conclusive results. In contrast, however, the aforementioned estimation method operates to sufficient accuracy over short periods time as a consequence of the drastic ratio between the masses of the Sun and any other bodies in the Solar System; it must be noted nonetheless that such an approximation fails when the planetesimal of interest approaches too near a major gravitational body other than the Sun or if the duration of simulation is not short enough. Furthermore, extrapolating from the premise of Chaos Theory [Regev 2006], predicting the future of these countless planetesimals drifting through empty space over the course of millions of years is onerous; so much so, minute changes in the initial conditions of the asteroid's orbital elements imply significant changes in the final outcomes over substantial time spans.

To the ends of overcoming such hurdles via the two-body approximation, the Swift and the Swift Visualization (SwiftVis) software packages (https://www.boulder.swri.edu/\~hal/swift.html) were executed, for the generation of graph plots given the simulation for data analysis over long-term integrations; the Swift software package was a courtesy of the Southwest Research Institute and can be briefly described as a package hosting a repertoire of algorithms for integration of mutual gravitational interactions between the planets, Sun, and test particles given the tests were not under the influence of gravity from neither themselves nor affecting other major bodies. To model the theoretical behavior of asteroid 1998 OH by inputting information pertaining to the planets as well as their orbits to reveal how they interact with theoretical particles, whose initial position and velocity vectors were also input into the program, the package applied the Regularized Mixed Variable Symplectic (RMVS) method to generate an integrator simulation. In addition, as the program evolved, one would be able to extract the orbital elements of all planets and particles at that certain timestep and also observe the last orbital elements of particles which are no longer active in the system. 
Since the OD program does not calculate the positions, velocities, nor orbital elements to exact precision, or to an accuracy even below about a minimum of $10^{-5}$ times the measurement, the generation of clone asteroids, each with slight manual perturbations to the position and velocity vectors calculated from using our full OD program on the mean $(\alpha, \delta)$ values from Table 6, was essentialized. For a random, independent selection of the six total vector components, one to four digits after the fifth digit ensuing the decimal point was designated a random change in their values by either addition or subtraction of one from each pick, thus creating a sample of test particles suitable for simulation implementation. Utilizing the above methodology, forty-five pseudo-asteroids were produced.

\FloatBarrier
\begin{figure}[h]
\begin{center}
\includegraphics[width=20pc]{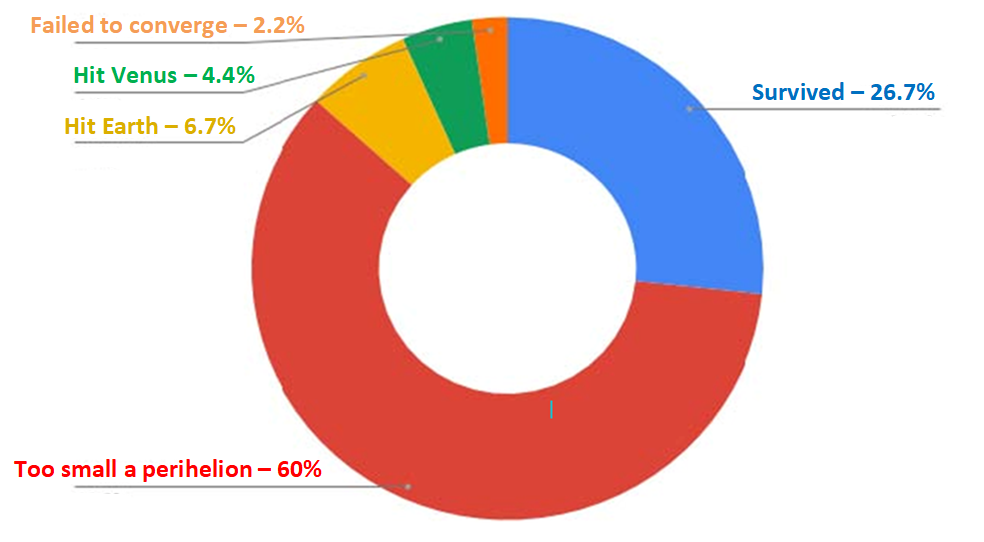}
\end{center}
\caption{\centering
Pie chart displaying the possible fates of our asteroid, 1998 OH.}
\end{figure}
\FloatBarrier

  \begin{table*}[t]
\tabularfont
\centering
\caption{\centering Osculating orbital elements obtained at the 50 Myr mark for analyzing clones (all measurements are given to 4 decimal places d.p.).}
\vspace{3mm}
  \begin{tabular}{c|c|c|c|c|c|c|c|c}
\hline
Identifier & a (AU) & e & i (rad) & $\Omega$ (rad) & $\omega$ (rad)& M (rad) & q (AU) & Q(AU) \\
\hline
-2 & 0.7223 & 0.0060 & 0.0708 & 1.1864 & 3.2585 & 1.0672 & - & - \\
-3 & 0.9986 & 0.0183 & 0.0335 & 2.3475 & 2.4469 & 4.4769 & - & - \\
-4 & 1.5237 & 0.0899 & 0.0421 & 1.7353 & 0.3009 & 0.4118 & - & - \\
-5 & 5.2027 & 0.0561 & 0.0340 & 2.0183 & 1.1237 & 3.2681 & - & - \\
-6 & 9.5283 & 0.0285 & 0.0229 & 1.3040 & 3.7878 & 5.1325 & - & - \\
-7 & 19.2071 & 0.0603 & 0.0098 & 1.8095 & 4.2612 & 1.3320 & - & - \\
-8 & 30.1233 & 0.0109 & 0.0184 & 2.0380 & 0.0735 & 2.9448 & - & - \\
\hline
1 & 1.8244 & 0.1227 & 0.3039 & 6.1780 & 3.9931 & 4.4020 & 1.6005 & 2.0482 \\
3 & 1.0991 & 0.8754 & 0.1146 & 6.0801 & 4.7670 & 2.9064 & 0.1370 & 2.0613 \\
8 & 0.9729 & 0.2237 & 0.9226 & 1.7902 & 1.2267 & 0.1458 & 0.7553 & 1.1905 \\
11 & 1.2916 & 0.9086 & 1.1213 & 4.8458 & 2.9299 & 4.5923 & 0.1180 & 2.4651 \\
\hline
\end{tabular}
\end{table*}

\begin{table*}[t]
\tabularfont
\centering
\caption{\centering Discarded Particles. L.P.E. signifies Last Planet Encountered (all values are 4 d.p.).}
\vspace{3mm}
  \begin{tabular}{c|c|c|c|c|c|c}
\hline
Identifier & Reason & L.P.E. & a (AU) & q (AU) & Q (AU) & i (deg) \\
\hline
2 & 1 & -2 & 1.0143 & 0.6774 & 1.3512 & 20.5623 \\
4 & 2 & -3 & 1.7794 & 0.0046 & 3.5542 & 37.1304 \\
5 & 2 & -3 & 2.4706 & 0.0045 & 4.9366 & 36.5190 \\
6 & 2 & -3 & 2.2493 & 0.0045 & 4.4941 &  29.5555 \\
7 & 2 & -3 & 2.0184 & 0.0045 & 4.0322 & 35.6852 \\
9 & 2 & -3 & 2.2637 & 0.0046 & 4.5228 & 31.7057 \\
10 & 2 & -3 & 2.1996 & 0.0045 & 4.3946 & 16.8626 \\
12 & 2 & -2 & 3.2350 & 0.0043 & 6.4657 & 23.7136 \\
13 & 3 & -3 & 0.6028 & 0.1632 & 1.0424 & 1.5643 \\
14 & 2 & -3 & 2.5115 & 0.0046 & 5.0184 & 45.8820 \\
15 & 2 & -2 & 2.0724 & 0.0044 & 4.1403 & 32.1470 \\
\hline
\end{tabular}
\end{table*}

Following the generation of $\vec{r}, \dot{\vec{r}}$ from the full OD program and the orbit determination of osculating orbital elements, the future of the clones were forecasted in a realistic chaotic environment, taking into account the major gravitational presence of planets (Tardioli {\em et al.}). The program ran a simulation of the next fifty million years (Myr) in thousand year time steps. Table 8 displays the orbital elements of planets alongside some of the particles still active at the end of the simulation and their perihelion (q) and aphelion (Q): Venus (-2), Earth (-3), Mars (-4), Jupiter (-5), Saturn (-6), Uranus (-7), and Neptune (-8). As seen in Fig. 12, out of the clones, 27 asteroids, or 60\% of the sample, eventually evolved to have too small a perihelion and collided with the Sun after encountering the Earth before their demise. This leads us to believe that 1998 OH will experience gravitational assist from the Earth after a close approach and consequently be given an acceleration in the direction of the Sun. Moreover, 3 of the asteroids, about 6.7\% of the total, were predicted to have an impact with Earth in the next 50 Myr, which could be potentially hazardous to humankind but based on its low probability of occurrence, will most likely not occur. Fig. 10 shows all eventual fates of our cloned asteroids, as predicted by Swift. Based on the simulation, 1998 OH is most likely to crash into the Sun in the next 50 Myr. Table 9 depicts the discarded particles with their identifier numbers and the reason behind their extinction concurrently: (1) collision with Earth, (2) too small a perihelion, and (3) collision with Venus.

\FloatBarrier
\begin{figure*}[t]
\centering
\begin{center}
\includegraphics[width=24pc]{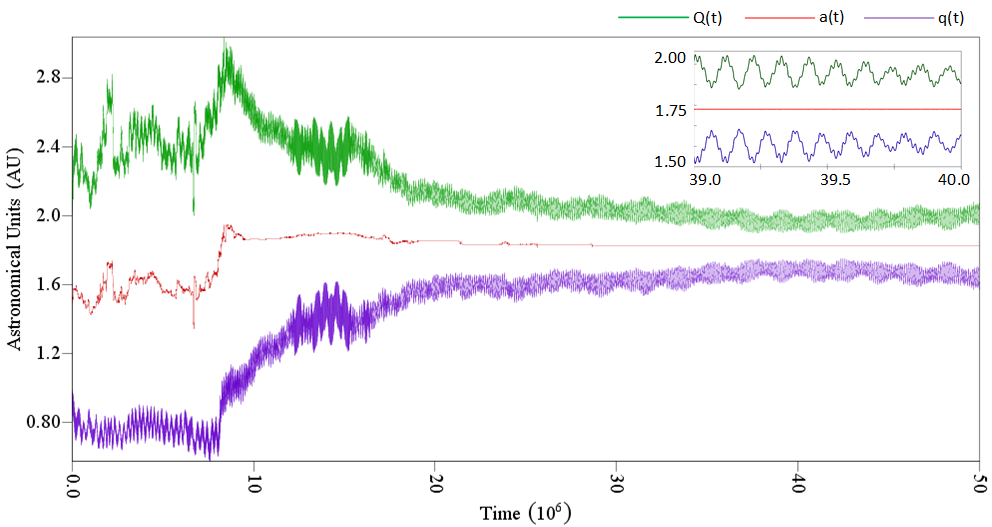}
\end{center}
\caption{\centering
\centering Particle 1 - Non-colliding, survived in the simulation.
}
\end{figure*}

\begin{figure*}[t]
\centering
\begin{center}
\includegraphics[width=24pc]{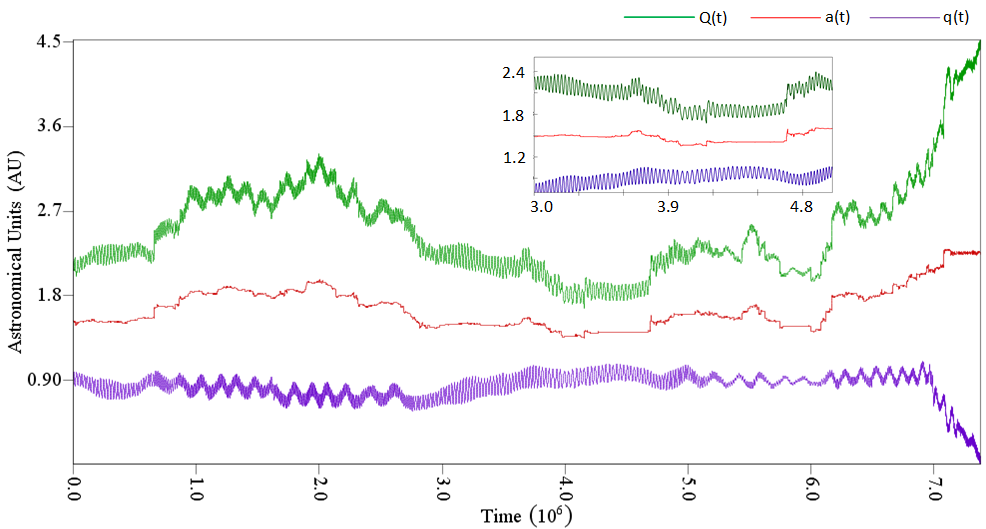}
\end{center}
\caption{\centering
Particle 6 - Collided into the Sun
}
\end{figure*}

\begin{figure*}[t]
\centering
\begin{center}
\includegraphics[width=24pc]{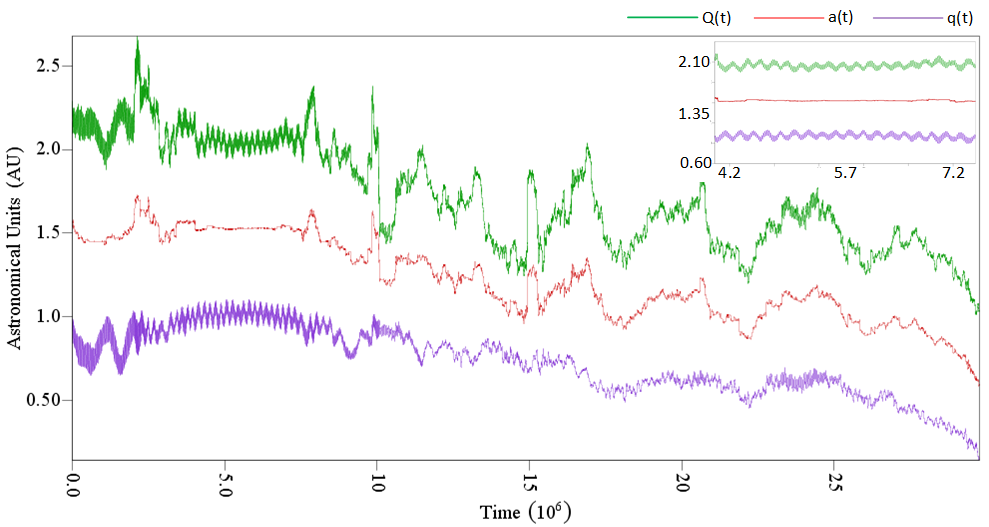}
\end{center}
\caption{\centering
Particle 13 - Collided with Venus}
\end{figure*}

\begin{figure*}[t]
\centering
\begin{center}
\includegraphics[width=27pc]{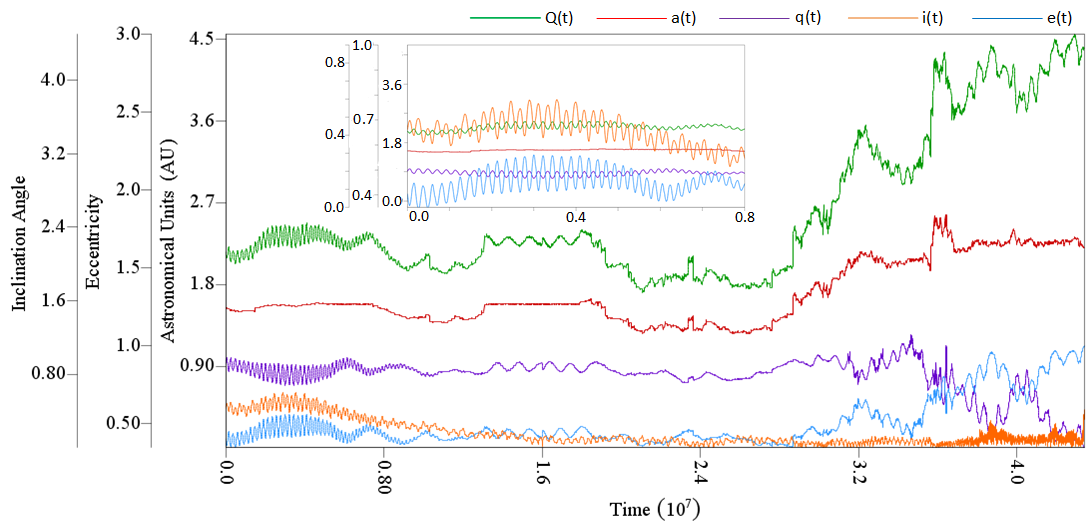}
\end{center}
\caption{\centering
 Particle 10 - Collided with the Sun. Initial stages of the clone's lifecycle exhibits noteworthy patterns}
\end{figure*}

\subsection{Data Analysis}
In addition, SwiftVision was operated to plot the semi-major axis (red), perihelion (purple), and aphelion (green) on the y-axis, whereas the eccentricity (blue) and angle of inclination (orange) were plotted for the same clone in Fig. 16. The x-axis represented the time in millions of years for all test cases. Four test particles that represented substantially different outcomes are displayed below for extended evaluations. The semi-major axis was measured to visualize the amount of energy the asteroids have whilst the eccentricity represented the trends in angular momentum.

For the first test case in Fig. 13, the simulation concluded that it survived without any collision with a massive body shown by the convergence of the aphelion and perihelion, leading to stability after some erratic fluctuation in the first fifteen Mya, likely due to a close encounter with another planet causing a drastic increase in the energy of the asteroid. Another notable characteristic for the graph of particle 1 is the sinusoidal pattern after the reaching of orbital stability. These slow alterations can be logically inferred and attributed to the regular, oscillatory gravitational effects of other planets. Considering its aphelion and perihelion became 2.0482 AU and 1.6005 AU respectively at the fifty Myr mark, the clone no longer had any planets crossing its orbit, explaining the sudden stabilization of the asteroid due to the reduced gravitational perturbations.

 As illustrated on the graph of Fig. 14 for particle 6, the simulation determined its fate to be a collision with the Sun as the divergence in the perihelion and aphelion increases rapidly at the seven million years mark. Consequently, the distance of the object from the Sun decreases rapidly and the object collides into it and diminishes. Moreover, the simulation determined the last planet of encounter as Earth, implying this was a case of gravitational slingshotting as can be seen from multiple points in the graph where the semi-major axis and aphelion drastically fluctuate despite the perihelion remaining somewhat constant, alluding to the possibility that the close encounters occurred near perihelion. At collision, the asteroid had an orbit with perihelion 0.0045 AU and aphelion 4.4941 AU, showing that upon collision, Venus, Earth and even Mars were all within the range of its orbit. This trend serves as an omen for its eventual death due to its incrementally unstable orbit throughout the simulation cycle. Particle 13, as displayed in Fig. 15, was of the few clones that collided with Venus after thirty Myr, as the erratically fluctuating decrease in the semi-major axis implies.
 
  Its last planet of encounter was Earth before the collision as its semi-major axis decreased rapidly and neared Venus's trajectory. Evidently, Particle 10 in Fig. 16 was found to exhibit too small a perihelion after the timescale of five Myr and impacted the Sun. The eccentricity and angle of inclination (in radians) are also plotted. A notable characteristic occurs over the course of eight million years from now when the semi-major axis remains approximately constant although the inclination increases slowly at the same rate as the eccentricity's decrease and a comparable correlation is exhibited by the perihelion and aphelion. Such indicates a resonant interaction, specifically speaking a Kozai-Lidov resonance mechanism (Kozai 1962; Lidov 1962) between the particle and a planet; extrapolating from the perihelion of 0.9 AU and an aphelion of 2.0 AU, this clone was most likely in orbital resonance with Earth or Mars as the only two major bodies crossing the pseudo-asteroid's orbit.
  
  \section{Conclusion and Outlook} \label{asteroid fly}
This paper proposes the most pragmatic technique and procedure possible to effectively determine the orbital trajectory of heliocentric objects, with the investigation's crux entailing the discussion on reliability of such methodology put forward with a test case-study. Astrometric and photometric data yielded with the telescopes returned near error-free results consistently and the observational data were submitted to the Minor Planet Center (MPC) for an updation to their database to support the cause of NEA (Near-Earth Asteroid) detection before coming into close proximity with the Earth. Such confidence in outcome stems from appropriate accuracy due to low discrepancy in results. Despite the consistent results and their substantiation of reliability, few findings deviated from the expected comparative values of Jet Propulsion Laboratory with considerable disparity. It is worth outlining the potential causes for such inconsistencies and likely improvements for further study.

 For some nights, the intermittent cloud coverage may have caused perturbations in the pattern of photons captured by the CCD camera, leading to inaccurate centroiding. Additionally, a physical malfunction in the telescope such as the probable misalignment of the reticule with the visual field could have lead to comparable implications. Dynamic atmospheric effects, in general, contributed to data acquisition in discordant ambiences. To account for such considerations, observations could have been conducted across different geographical vantage points to enable for parallax correction [Meeus 1998] to minimize the effects of stellar aberration. As discovered in the ephemeris generation section, an obsolete database is a possible ramification and in the future, differential correction can be used to improve accuracy in findings.

Gauss's Method for determining position and velocity vectors proved laborious for the asteroid; depending on the observations used, vectors would either converge to a reasonable value or fail to do so. We encountered the latter when attempting to analyze the trailing data from observations in late July, 2019. Although output vectors were deemed more accurate than ephemeris values, they were expected to be somewhat consistent; in contrast, observations from late July were producing large percent errors likely due to 1998 OH's rapidly increasing declination as it migrated above the ecliptic in its prograde orbit. In order to counteract such abnormalities, the conducting of observations across the 1998 OH's orbital period (approximately 699 days) is recommended to achieve a more insightful determination of the asteroid throughout its revolution in equally spaced time intervals. 

To further elaborate, the Method of Gauss was given preference over another well-established orbital determination technique known as Method of Laplace. Although both work to appropriate standards in relatively short periods of time, Gauss is more accurate when the range vectors are relatively small whereas Laplace functions optimally at low magnitudes of the position vectors. Since the 1998 OH was a NEO, Gauss's Method adhered to the indicated criterion more appropriately, resulting in close approximate solutions over large recursions; the such provided acceptable accuracy for data, but increasing the number of used Taylor expansion terms would allow the calculation to be even more exact for the elements. It would also be insightful to apply the same test case-study using Laplace's Method to enable comparison between the reliability of each method beyond speculation using achieved results.

The effectiveness of the proposed strategy can also be showcased with asteroid impact risk evaluation run by the Swift simulation output. Based on which, 1998 OH has a larger probability of death by hitting the Sun, with more than half of the test particles suffering the identical fate as demonstrated by our improved methodology. However, 40\% of our other test particles did not end up like so; hence it is considerably probable 1998 OH would not eventually die from too small a perihelion.

Key limitation in Swift simulations was the insufficient amount of storage, execution time, and clock-speed of our computers which prevented the simulation of a larger pseudo-asteroid sample to generate more representative values. Nonetheless, the migration of asteroids in Kozai resonance mechanisms tend to gravitate towards unstable ones, implicating the greater chance of asteroids' chaotic orbital path deviation and henceforth its manifold possibilities of demise, primarily due to the high eccentricity causing too small a perihelion and its eventual collision with the Sun.

\appendix

\section{Formulaic Application Of The Least Squares Plate Reduction Algorithm}

Given: (x, y) for N stars and the asteroid from the plate alongside the ($\alpha$, $\delta$) of the reference stars, find the ($\alpha$, $\delta$) with minimum error. Cartesian and celestial coordinates have differing origins, orientations, and scale of magnitudes, hence conduct linear transformations by the relations:
\begin{equation}
    \alpha = b_1+a_{11}x+a_{12}y
\end{equation}
\begin{equation}
    \delta  = b_2+a_{21}x+a_{22}y
\end{equation}
where the plate coefficients, b$_{1}$ and b$_{2}$ represent offsets whereas a$_{11}$, a$_{12}$, a$_{21}$ and a$_{22}$ rotate and scale the coordinate system linearly. Afterwards, the linear regression must be applied by the following formulae:
\begin{equation}
    d_{i\alpha} = \alpha_{i}-(b_1+a_{11}x_i+a_{12}y_i)
\end{equation}
\begin{equation}
    d_{i\delta} = \delta_{i}-(b_2+a_{21}x_i+a_{22}y_i)
\end{equation}
Define the "chi-squared" statistic and set the partial derivatives equal zero for minimization of error:
\begin{equation}
    \chi_\alpha^{2} = \sum_{i=1}^{N} [d_{i\alpha}]^{2} \quad \textrm{and} \quad \chi_\delta^{2} = \sum_{i=1}^{N} [d_{i\delta}]^{2}
\end{equation}
\begin{equation}
    \frac{\partial \chi_\alpha^{2}}{\partial b_1} = 0 \quad , \quad \frac{\partial \chi_\alpha^{2}}{\partial a_{11}} = 0 \quad , \quad \frac{\partial \chi_\alpha^{2}}{\partial a_{12}} = 0
\end{equation}
\begin{equation}
    \frac{\partial \chi_\delta^{2}}{\partial b_2} = 0 \quad , \quad \frac{\partial \chi_\delta^{2}}{\partial a_{21}} = 0 \quad , \quad \frac{\partial \chi_\delta^{2}}{\partial a_{22}} = 0
\end{equation}
The procedure for deriving the three unknowns and their equations are tantamount for the right ascension and declination. Taking the right ascension scenario:
\begin{equation} \label{eq1}
\begin{split}
    \frac{\partial \chi_\alpha^{2}}{\partial b_1} & = \sum 2(d_{i\alpha}) \frac{\partial d_{i\alpha}}{\partial b_1} \\
    &\Rightarrow -2\sum[\alpha_i-(b_1+a_{11}x_i+a_{12}y_i)]=0 \\
    &\Rightarrow \sum{a_i} = b_1N+a_{11}\sum x_i+a_{12}\sum y_i
\end{split}
\end{equation}
\begin{equation}
    \begin{split}
        \frac{\partial \chi_\alpha^{2}}{\partial a_{11}} & = \sum 2(d_{i\alpha}) \frac{\partial d_{i\alpha}}{\partial a_{11}} \\
        &\Rightarrow 2\sum x_i[\alpha_i-(b_1+a_{11}x_i+a_{12}y_i)]=0 \\
        &\Rightarrow \sum {a_ix_i} = b_1\sum x_i + a_{11}\sum x_{i}^{2} + a_{12}\sum x_iy_i
    \end{split}
\end{equation}
\begin{equation}
    \begin{split}
        \frac{\partial \chi_\alpha^{2}}{\partial a_{12}} & = \sum 2(d_{i\alpha}) \frac{\partial d_{i\alpha}}{\partial a_{12}} \\
        &\Rightarrow 2\sum y_i[\alpha_i-(b_1+a_{11}x_i+a_{12}y_i)]=0 \\
        &\Rightarrow \sum {a_iy_i} = b_1\sum y_i + a_{11}\sum x_iy_i +a_{12}\sum y_i^{2}
    \end{split}
\end{equation}
The preceding can be organized into a matrix solvable by the techniques of Cramer's Law. Applying the identical process for the declination leads to a comparable matrix. Note the matrices are symmetrical about the diagonal:
\begin{equation}
\begin{bmatrix}
    \sum{\alpha_i} \\
    \sum {\alpha_ix_i} \\
    \sum {\alpha_iy_i}
  \end{bmatrix}
=
\begin{bmatrix}
  N & \sum x_i & \sum y_i \\
  \sum x_i & \sum x_i^{2} & \sum x_iy_i \\
  \sum y_i & \sum x_iy_i & \sum y_i^{2}
\end{bmatrix}
\begin{bmatrix}
    b_1 \\
    a_{11} \\
    a_{12}
\end{bmatrix}
\end{equation}
\begin{equation}
\begin{bmatrix}
    \sum{\delta_i} \\
    \sum {\delta_ix_i} \\
    \sum {\delta_iy_i}
  \end{bmatrix}
=
\begin{bmatrix}
  N & \sum x_i & \sum y_i \\
  \sum x_i & \sum x_i^{2} & \sum x_iy_i \\
  \sum y_i & \sum x_iy_i & \sum y_i^{2}
\end{bmatrix}
\begin{bmatrix}
    b_2 \\
    a_{21} \\
    a_{22}
\end{bmatrix}
\end{equation}
Succeeding the manual computation of the right ascension and declination, the residuals are calculated by (52) and (53) whilst their uncertainties derived from:
\begin{equation}
    \sigma_\alpha^{2} \Rightarrow \frac{\chi_\alpha^{2}}{N-3} = \frac{1}{N-3} \sum_{i} d_{i\alpha}^{2} 
\end{equation}
\begin{equation}
    \sigma_\delta^{2} \Rightarrow \frac{\chi_\delta^{2}}{N-3} = \frac{1}{N-3} \sum_{i} d_{i\delta}^{2}
\end{equation}
Note the "N-3" emerges from the requirement that a minimum of three reference stars must be sampled for efficacious outcomes.

\section{Sample Code For Plotting The Orbital Trajectories}
\begin{lstlisting}[language=Python]
from math import *
from mpl_toolkits.mplot3d import Axes3D
import numpy as np
import matplotlib.pyplot as plt

#The OD library of functions
import odlib as od

fig = plt.figure()
ax = fig.add_subplot(111, 
projection='3d')
plt.title('Orbit of asteroid 12538
(1998 OH)')
ax.set_xlabel('x (AU)')
ax.set_ylabel('y (AU)')
ax.set_zlabel('z (AU)')

#list of time ranges
trange = np.arange(0, 2000, 1) 

a= 1.5135 #Semi-Major Axis

e= 0.396045 #Eccentricity

i= 24.288149 #Inclination

w= 320.920565 #Argument of Perihelion

#Crude Mean Anomaly from JPL
M_2= 43.883072*(pi/180) 

#Longitude of Ascending Node
omega= 221.049404  

#Julian date for time of observation
t2 = 2458668.716975

k = 0.01720209847

#date for which mean anomaly is desired
t = 2458698.784988426 

#Implementation of ephemeris generation
def n(a): #equation for n
    return(sqrt(1/(a)**3))

def T(a, t2, M_2,k): #equation for T
    return(t2 - (M_2/(n(a)*k)))

def M(a,t2,t,k,M_2): #equation for M
    return(n(a)*k*(t-T(a,t2,M_2,k)))

#Newtons method used to guess E
def function(E,a,t2,t,k,M_2):
    e=0.396045
    
    N=M(a,t2,t,k,M_2)
    return E-e*sin(E)-N

def derivative(E,a,t2,t,k,M_2):
    e=0.396045
    return 1-e*cos(E)

def newton_method(xi,a,t2,t,k,M_2):
    while (abs(function(xi,a,t2,t,k,M_2))
    >0.00000001):
        xi=xi-function(xi,a,t2,t,k,M_2)
        
        derivative(xi,a,t2,t,k,M_2)
    return xi

#equations for r vector
r= np.array([a*cos(newton_method(0,a,t2,t,
k,M_2))-a*e,a*sqrt(1-e**2)*
sin(newton_method
(0,a,t2,t,k,M_2)), 0])

#rotate r clockwise in sequence
ecliptic = od.rotation(*r,-omega,-i,-w)

x = []
y = []
z = []

#Calculate r for all time ranges
for j in trange: 
    r = np.array([a*cos(newton_method
    (0,a,t2,j,k,M_2))-a*e, a
    *sqrt(1-e**2)*sin(newton_method
    (0,a,t2,j,k,M_2)), 0])
    ecliptic=od.rotation(*r,-omega,
    -i,-w)
    
    x.append(ecliptic[0])
    y.append(ecliptic[1])
    z.append(ecliptic[2])

ax.scatter3D(x, y, z)

#the Sun as a reference benchmark
ax.plot([0],[0],[0],'oy')

plt.show()

\end{lstlisting}

\vspace{3mm}
\section*{Acknowledgements}
This research has been fully funded by the gracious grant of the Summer Science Program (SSP) 2019 conducted at Sommers-Bausch Observatory, University of Colorado Boulder, United States of America. We express sincere gratitude to our colleagues Patrick Liu and Zoe Curewitz for their technical advice relating to the asteroid plot. A special thanks goes to Dr. Katherine Kretke and Dr. Raluca Rufu for their contribution to our asteroid simulations.

\end{document}